\documentclass[journal]{vgtc}                     

\DeclareGraphicsExtensions{.png,PNG,.pdf,.PDF}

\onlineid{1595}

\vgtccategory{Research}

\DeclareMathAlphabet{\mathcal}{OMS}{cmsy}{m}{n}

\def \etal {{\emph{et al}.\thinspace}}

\def \eg {{\emph{e.g}.\thinspace}}
\def \ie {{\emph{i.e}.\thinspace}}

\title{Dynamic Color Assignment for Hierarchical Data}

\author{%
  \authororcid{Jiashu Chen*}{0009-0002-0412-7212},
  \authororcid{Weikai Yang*}{0000-0002-6520-1642},
Zelin Jia,
Lanxi Xiao,
and \authororcid{Shixia Liu}{0000-0003-4499-6420}
}

\authorfooter{
  \item
  	J. Chen, W. Yang, Z. Jia, L. Xiao, S. Liu are with the School
of Software, BNRist, Tsinghua University. J. Chen and W. Yang are joint first authors. S. Liu is the corresponding author.
  	E-mail: \{\{cjs22,yangwk21,jiazl22, xlq20\}@mails., shixia@\}\ tsinghua.edu.cn
}

\abstract{Assigning discriminable and harmonic colors to samples according to their class labels and spatial distribution can generate attractive visualizations and facilitate data exploration.
However, as the number of classes increases, it is challenging to generate a high-quality color assignment result that accommodates all classes simultaneously.
A practical solution is to organize classes into a hierarchy and then dynamically assign colors during exploration.
However, existing color assignment methods fall short in generating high-quality color assignment results and dynamically aligning them with hierarchical structures.
To address this issue, we develop a dynamic color assignment method for hierarchical data, which is formulated as a multi-objective optimization problem. 
This method simultaneously considers color discriminability, color harmony, and spatial distribution at each hierarchical level.
By using the colors of parent classes to guide the color assignment of their child classes, our method further promotes both consistency and clarity across hierarchical levels.
We demonstrate the effectiveness of our method in generating dynamic color assignment results with quantitative experiments and a user study.\looseness=-1
}

\keywords{Color assignment, Hierarchical Visualization, Discriminability, Harmony.}

\teaser{
  \centering
  \includegraphics[width=\linewidth]{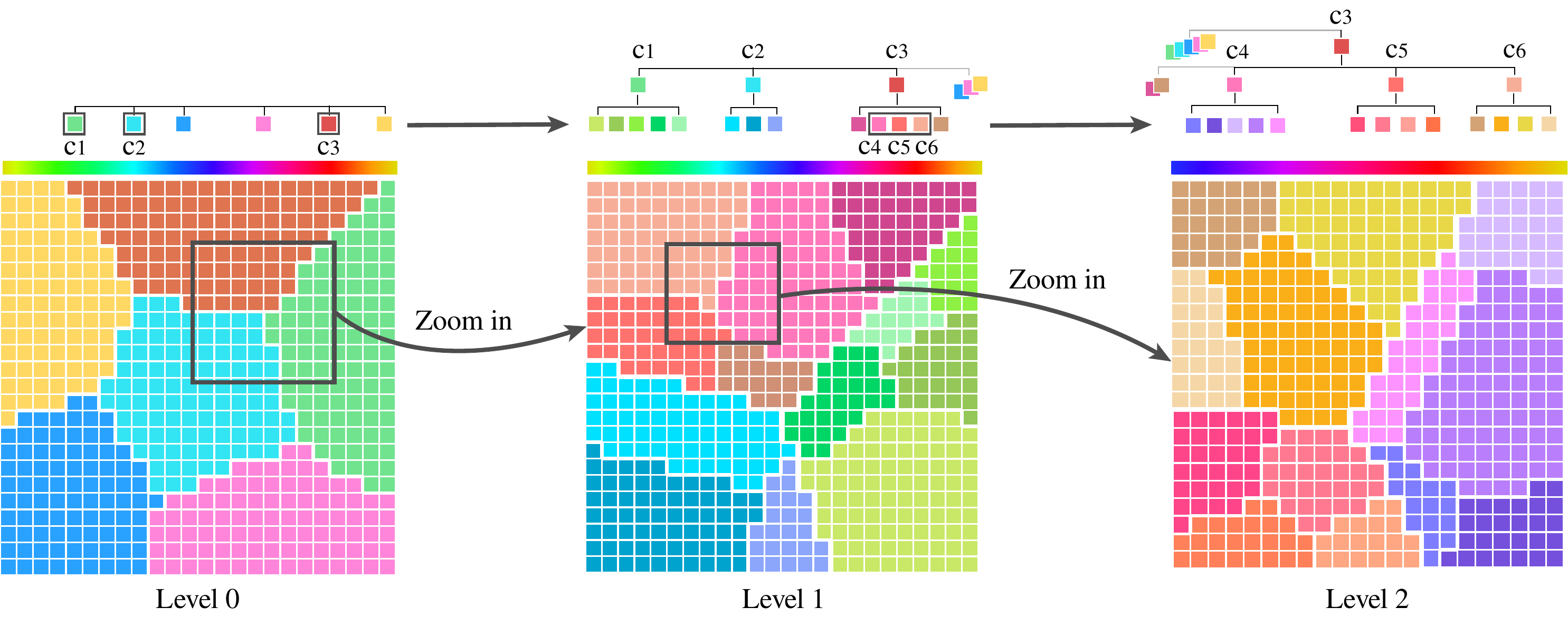}
  \caption{Based on user exploration, our method dynamically selects the color range and assigns colors to classes within the range, which ensures high discriminability and harmony at each level and maintains consistency across different levels.
  }
  \label{fig:teaser}
}

\graphicspath{{figs/}{figures/}{pictures/}{images/}{./}} 

\usepackage{multirow}
\usepackage{color}
\usepackage{lipsum}                    
\usepackage{amssymb,amsmath}
\usepackage{wrapfig}
\usepackage{booktabs}
\usepackage{makecell}
\usepackage{paralist}
\usepackage{algorithm}
\usepackage{algpseudocode}
\usepackage{dblfloatfix}

\newcommand{\vica}[1]{\textcolor{black}{#1}}

\newcommand{\crule}[1]{\textcolor[RGB]{#1}{\rule{0.22cm}{0.22cm}}\:\!}

\usepackage{mathptmx}                  

\begin{document}


\firstsection{Introduction}

\maketitle
\fontsize{9}{9} 
Assigning colors to samples according to their class labels and spatial distribution is a common practice in data analysis~\cite{yang2022diagnosing,yang2024foundation,liu2024visualization}.
A high-quality color assignment should be discriminable and harmonic to ensure clarity and attractiveness in visualizations~\cite{Gramazio2017colorgorical,liu2022image}.
However, as the number of classes increases, it is challenging to select a large number of colors that are easily distinguishable from each other but also harmonize together ~\cite{tseng2023measuring,fang2016categorical}.
A practical solution is to organize classes into a hierarchical structure and then dynamically assign consistent colors across hierarchical levels. 
This solution not only enhances scalability by reducing the requirement for a large number of distinct colors in a static visualization, but also alleviates cognitive load during data exploration.
However, existing color assignment methods do not fully support this solution.
Some methods generate color assignment results independently for each hierarchical level, which cannot maintain color consistency across levels~\cite{lu2021palettailor,fang2016categorical}. 
This will disrupt the user's mental map during exploration.
Other methods apply strict color constraints to achieve consistency but sacrifice discriminability and harmony at each level~\cite{tennekes2014treecolor,fua1999hierarchical}.
These limitations highlight the need for a dynamic color assignment method that is capable of maintaining color discriminability and harmony at each level and ensures consistency across different levels during exploration.
\begin{table*}[!b]
\caption{The comparison between several representative color assignment methods and our method.}
\setlength{\tabcolsep}{1.1em}
\begin{tabular}{cccccc}
   \toprule
   Methods & Discrimination & Harmony & Spatial Distribution & Alignment with Hierarchy & Dynamic Assignment\\
   \midrule
   Palettailor~\cite{lu2021palettailor} & \checkmark & - & \checkmark & - & -\\
   Color Crafting\cite{smart2019color} & \checkmark & \checkmark & - &  -& - \\
   Tree Colors\cite{tennekes2014treecolor} & \checkmark & - & -  & \checkmark & -\\
   Cuttlefish\cite{waldin2019cuttlefish} & \checkmark & - & -  & \checkmark & \checkmark\\
   Ours & \checkmark & \checkmark & \checkmark& \checkmark & \checkmark\\
   \bottomrule
\end{tabular}
\label{tab:methods}
\end{table*}

To determine the design requirements for developing such a dynamic color assignment method, we first conduct interviews with six experts specializing in Information Design in a School of Arts.
The findings indicate that the most important goal is to ensure discriminability, which facilitates identifying class labels of data samples.
Following this, harmony is identified as the second most important factor, critical to producing visually attractive results that engage users.
They also point out that considering spatial distribution can further improve discriminability and harmony and thus facilitate data analysis. 
For example, class boundaries can be made clearer by increasing discriminability between adjacent classes~\cite{lu2021palettailor}.
Accordingly, we formulate the color assignment as a multi-objective optimization problem with suggested priorities among these objectives~\cite{mahapatra2020multi}, and apply the most advanced theories to quantify each objective.
Next, to solve this complex optimization problem, we employ simulated annealing for its high flexibility in handling multiple objectives, and combine it with the continuation method to sequentially incorporate discriminability, harmony, and spatial distribution during the optimization process.
This accelerates convergence to a better solution by guiding the optimization process towards more promising regions in the solution space~\cite{bengio2009curriculum}.
To generate dynamic color assignment results based on user exploration and align them with the hierarchical structures within datasets, the colors of the parent classes are used to guide the color assignment of their child classes.
This is achieved by dynamically selecting appropriate color ranges for child classes based on the colors of their parent classes, and then optimizing the color assignment result within the selected color range.
As shown in Fig.~\ref{fig:teaser}, our method achieves color consistency across levels in hierarchical grid visualizations.

Quantitative experiments show that compared to state-of-the-art methods, our method performs best in ensuring discriminability and aligning with hierarchical structures, while still offering comparable levels of harmony.
A user study with 20 experts further confirms that our method generates high-quality color assignment results that are closely aligned with user preference.

The main contributions of our work include:\looseness=-1
\begin{compactitem}

\item\noindent A color assignment method that achieves better discriminability and harmony.
\item\noindent A dynamic color range selection method in which the colors of the parent classes guide the color assignment of their child classes.
\item\noindent An open-source implementation of the proposed color assignment method in both C++ and JavaScript, available at \href{https://github.com/thu-vis/Dynamic-Color}{https://github.com/thu-vis/Dynamic-Color}.
\end{compactitem}

\section{Related Work}

Existing color assignment methods can be classified into two categories based on how they organize classes in a dataset: flat color assignment and hierarchical color assignment.

\textbf{Flat color assignment} methods assign colors to all classes without considering their hierarchical relationships.
In flat color assignment, ensuring discriminability between colors of different classes is a fundamental requirement and consistently draws research attention over the years~\cite{healey1996choosing,nardini2021automatic,maxwell2000visualizing,tufte1991envisioning, zheng2022image}.
As a pioneering study, Healey~\cite{healey1996choosing} proposed a rule-based method to select discriminable colors on the hue wheel that maximize perceptual differences and name differences.
Later studies improve discriminability by incorporating more advanced theory in quantifying perceptual differences and/or name differences~\cite{fang2016categorical,fang2016categorical,setlur2015linguistic}.
For example, Fang~\etal~\cite{fang2016categorical} calculated perceptual differences using CIEDE2000~\cite{sharma2005ciede2000}, which improves perceptual uniformity and achieves better alignment with human perception.
Setlur~\etal~\cite{setlur2015linguistic} utilized the name distance proposed by Heer and Stone~\cite{heer2012color}, which includes 153 popular color names and their color-name associations.
This offers a more precise way to measure name differences.

In addition to color discriminability, it is also important to generate harmonic and visually appealing color assignment results~\cite{cohen2006color,wang2008color,kita2016aesthetic,Gramazio2017colorgorical,smart2019color,yuan2021infocolorizer,wu2023adaptive, lin2022C3,liu2022image}.
For example, Cohen-Or~\etal~\cite{cohen2006color} introduced a color harmonization method by aligning colors with Matsuda's established harmonic templates~\cite{matsuda1995color}.
Color Crafting~\cite{smart2019color} summarizes the templates of designer-crafted color assignment results in the color space and then generates more color assignment results that mimic designer practices.

Color discriminability and harmony can be further improved by considering the spatial distribution of the visualized data~\cite{Lee2013percept,chen2014visual,wang2018optimizing,lu2021palettailor,lu2023interactive}.
Wang~\etal~\cite{wang2018optimizing} considered the color discriminability between neighboring points and their contrasts to the background in a scatterplot.
They then employed a genetic algorithm to find the best color assignment result from a set of pre-defined colors that maximizes discriminability.
Palettailor~\cite{lu2021palettailor} advances this method by simultaneously adjusting and assigning colors during the optimization process.

Although recent efforts in flat color assignment have achieved certain success in producing discriminable and harmonic colors, they struggle in real-world scenarios where the number of classes can reach hundreds or even thousands. 
To address the scalability issue in flat color assignment methods, researchers have developed several \textbf{hierarchical color assignment} methods. 
Early efforts focus on generating coherent colors for a static visualization, where all colors are visible at the same time~\cite{fua1999hierarchical,delon2005automatic,tennekes2014treecolor}.
For example, Fua~\etal~\cite{fua1999hierarchical} proposed a proximity-based coloring method for hierarchical parallel coordinates.
It recursively assigns colors to child classes within a range centered on the color of their parent class.
This range becomes progressively narrower at each level, which ensures that the colors of classes of the same parent are more similar to each other than those of different parents.
Similarly, Tree Colors~\cite{tennekes2014treecolor} divides the hue wheel into several ranges and assigns each to different branches of a tree.
As the level increases, the hue range for each branch narrows down, while the saturation increases.
This method generates a diverse but consistent color assignment result across the hierarchy.
However, as the number of classes increases exponentially with the levels, these methods still suffer from scalability issues regarding color discriminability.
To address this issue, later efforts adopt dynamic color assignment that only assigns colors to visible data during exploration~\cite{waldin2016chameleon,waldin2019cuttlefish}.
These methods better exploit the color space and thus improve the overall quality of color assignment results. 
For example, Chameleon~\cite{waldin2016chameleon} uses a force-based method to dynamically adjust color ranges on the hue wheel.
This method aims to keep the ranges close to their original position while reducing overlaps between adjacent ranges.
However, the force-based method does not prevent color overlaps between child classes of different parents.
Cuttlefish~\cite{waldin2019cuttlefish} extends it to eliminate overlaps by imposing hard constraints.
It allows a larger shift on the hue wheel to ensure distinct color ranges.

While these hierarchical color assignment methods succeed in maintaining color consistency between parent classes and child classes, they still face two issues.
First, these methods do not simultaneously consider discriminability, harmony, and spatial distribution to produce a high-quality color assignment result at each level.
Second, the strict constraints between the colors of parent classes and child classes often lead to insufficient discriminability, especially among the child classes of the same parent.
In comparison, we achieve a well-balanced integration of discriminability, harmony, and spatial distribution in our optimization process by formulating color assignment as a multi-objective optimization problem with suggested priorities among these tasks.
We also developed an improved color range selection method to enhance discriminability without sacrificing color consistency across class hierarchies.
\vica{The detailed comparison between representative color assignment methods and our method is summarized in Table~\ref{tab:methods}.}

\section{Requirement Analysis}
\label{sec:requirement}

\begin{figure}[t]
    \centering
    \includegraphics[width=\linewidth]{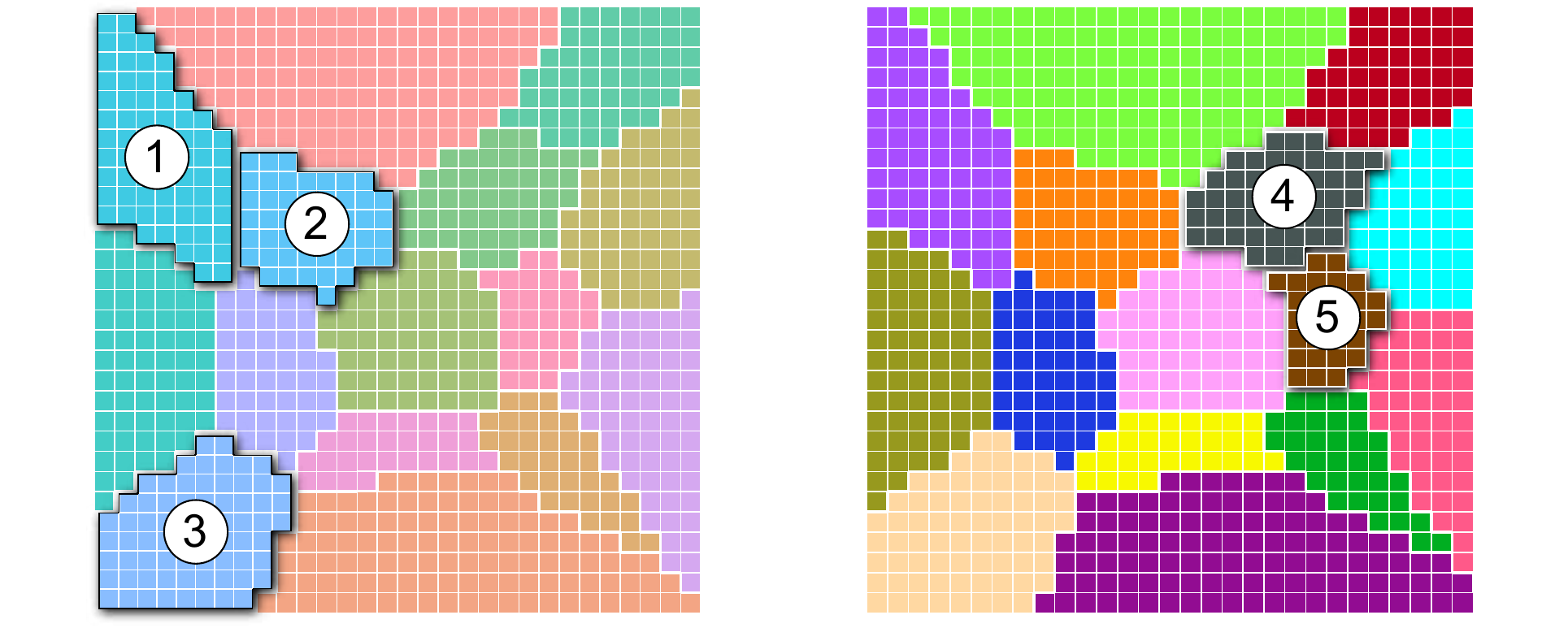}
    \put(-210,-10){(a) Cuttlefish}
\put(-90,-10){(b) Palettailor}
     \caption{Results generated by existing color assignment methods.}
    \vspace{-3mm}
    \label{fig:food-example}
\end{figure}

We worked closely with six experts (E1-E6) during the development of the dynamic color assignment method.
All of them major in Information Design in a School of Arts and have more than 5 years of experience in designing colors for visualizations and/or user interfaces.
E1 is the co-author of this paper, while E2-E6 are not.
We conducted six semi-structured interviews with each expert to collect the requirements for dynamic color assignment.
Initially, we shared the results of existing color assignment methods (\eg, Fig.~\ref{fig:food-example}) with the experts.
Then, they were asked to evaluate these results, highlighting both strengths and weaknesses.
They were also recommended to modify the colors to express their preferences.
Finally, we collected their advice on generating high-quality color assignment results and the factors that warrant particular attention.
Each interview lasted between 35 and 45 minutes.
In addition to these interviews, we also engaged in biweekly free-form discussions to showcase our color assignment results and promptly collect their feedback.

Based on the six semi-structured interviews, the biweekly discussions, and the literature review, we summarized four design requirements for dynamic color assignment.

\begin{figure*}[t]
    \centering
    \includegraphics[width=\linewidth]{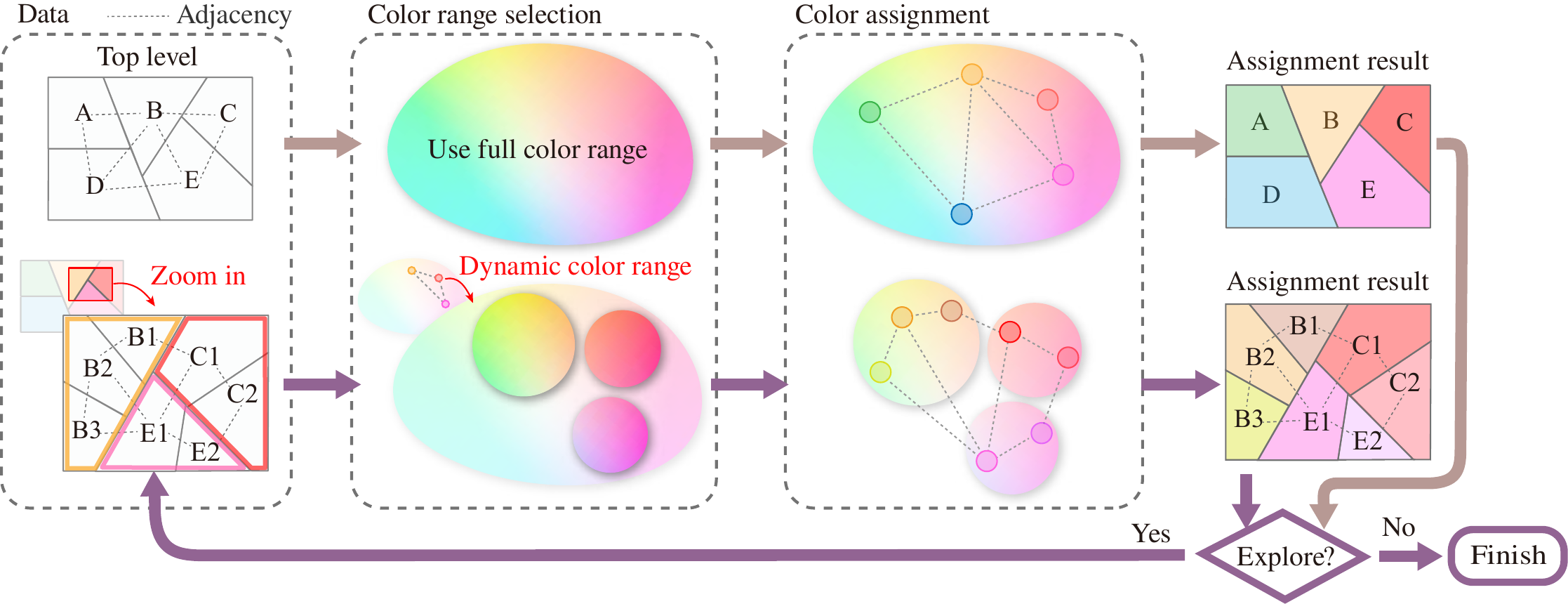}
    \caption{Method overview. Given data, the color range selection module selects an appropriate range of colors. Then, the color assignment module generates a discriminable and harmonic color assignment result within that range.}
    \label{fig:overview}
\end{figure*}

\vspace{1.0mm}
\noindent\textbf{Ensure color discriminability}.
All the experts agreed that color discriminability is the most important factor and should be considered first.
E1 commented that a minimum threshold of color differences is required to quickly identify different class labels.
This is also reflected in several previous research~\cite{stone2014engineering, brychtova2017effect, Gramazio2017colorgorical, lu2021palettailor}.
When examining color assignment results generated by existing methods, E2 and E4 pointed out that Color Crafting~\cite{smart2019color} and Cuttlefish~\cite{waldin2019cuttlefish} failed to achieve good color discriminability when the number of colors exceeded 10.
As shown in Fig.~\ref{fig:food-example}(a), the color assignment result generated by 
Cuttlefish results in three colors \crule{63,202,227}\crule{94,197,248}\crule{137,189,255} that are not sufficiently distinguishable from each other.
This is because this method mainly considers the difference in the hue channel, which limits its selection range.
A better color discriminability can be achieved by modifying their saturation and luminance (\crule{63,202,227}\crule{94,197,248}\crule{137,189,255} \textit{vs.} \crule{113,224,255}\crule{179,231,255}\crule{128,179,237}).
Therefore, it is necessary to simultaneously consider differences in hue, saturation, and luminance when generating color assignment results.

\vspace{1.0mm}
\noindent\textbf{Enhance color harmony}.
All the experts pointed out that they would also consider color harmony when choosing colors in their designs.
In this process, they would avoid using strongly disliked colors, such as DarkSlateGray (\crule{71,85,84}) and SaddleBrown (\crule{125,68,2}) in Fig.~\ref{fig:food-example}(b).
The strategy of excluding strongly disliked color ranges has also been widely employed in existing color assignment methods~\cite{lu2021palettailor,Gramazio2017colorgorical}.
When discussing how to improve color harmony, three experts pointed out that according to harmonic template theory~\cite{matsuda1995color,cohen2006color}, harmonic colors usually conform to specific geometric patterns in the color space, such as the hue wheel.
Four experts also noted that, in addition to the hue wheel, the balance between saturation and luminance is also crucial to color harmony.

\vspace{1.0mm}
\noindent\textbf{Consider spatial distribution}.
Our experts also highlighted the importance of considering spatial distribution in generating color assignment results.
Taking spatial distribution into account not only enhances data analysis~\cite{Lee2013percept, wang2018optimizing, yang2023survey} but also improves the aesthetic appeal of the results~\cite{ou2011additivity, li2023color,jiang2024region}.
E3 and E6 emphasized that the assigned colors of two spatially adjacent classes significantly affect perception and thus deserve careful consideration.
For example, enhancing the color discriminability between spatially adjacent classes can make the class boundaries clearer and aid in identifying different classes.
Moreover, maintaining color harmony between adjacent classes can produce more visually pleasing results.
E5 also noted that assigning similar colors to similar classes would facilitate data understanding and exploration.

\vspace{1.0mm}
\noindent\textbf{Align with hierarchical structures}.
As rich hierarchies are ubiquitous in datasets~\cite{yang2021interactive,chen2024enhancing,chen2024visualization},
four experts also acknowledged that the generated color assignment results should accurately reflect the hierarchical structures. 
E1 said, ``It is common practice to use similar colors to encode a parent class and its corresponding child classes, which facilitates the identification of hierarchical relationships and keeps the user's mental map throughout the zooming process.''
E4 further emphasized that to avoid misinterpretation of hierarchical relationships, the color differences between child classes of the same parent class should be smaller than those between child classes of different parent classes.

\section{Dynamic Color Assignment}
\label{sec:assignment}
\subsection{Method Overview}
Driven by the identified requirements, we propose a dynamic color assignment method that aligns well with the class hierarchy across levels.
At each level, our method simultaneously considers discriminability, harmony, and spatial distribution.
As shown in Fig.~\ref{fig:overview}, our method consists of two modules: \textbf{color range selection} and \textbf{color assignment}.
The color range selection module selects an appropriate color range to ensure consistency across hierarchical levels, and the color assignment module generates high-quality color assignment results within the selected color range.
Specifically, when assigning colors for top-level classes, the color range selection module selects the full color range that allows greater flexibility for generating color assignment results.
When users focus on a specific region for closer examination, this module selects the appropriate color range for the child classes based on the colors of their parent classes.
This ensures color consistency and provides a coherent exploration experience that adapts to user interactions.
Based on this exploration process, we will first introduce how to generate high-quality color assignment results at each level and then describe how to ensure consistency across levels using dynamic color range selection.

\subsection{Color Assignment}
\label{subsec:assignment}
The expert interviews reveal that generating high-quality color assignment results requires optimizing multiple goals, including discriminability, harmony, and spatial distribution.
However, due to the conflicting nature of these goals, it is impossible to maximize all of them simultaneously.
For example, the color assignment result with the best discriminability will contain colors with extremely high luminance, leading to lower harmony.
In practice, there are multiple Pareto-optimal solutions that cannot be enhanced in one goal without compromising another.
Existing efforts require users to adjust the weighting parameters of different goals to explore different Pareto-optimal solutions.
However, it brings an extra burden for users to fine-tune the weighting parameters and determine a better one.
In the requirement analysis, we have identified a priority order of \textit{discriminability} > \textit{harmony} > \textit{spatial distribution}.
Therefore, we utilized a priority-specific Pareto-optimal strategy, which is effective in identifying the Pareto-optimal solution that satisfies the priority order~\cite{mahapatra2020multi}.
This is achieved by guaranteeing that goals with higher priorities have higher objective values.
Specifically, given a set of colors $c_1,c_2,\ldots,c_m$, the optimization problem is formulated as:

\begin{equation}
\begin{split}
\max_{c_1,c_2,\ldots,c_m}\quad E_\text{D}&+ \alpha E_\text{H}+ \beta E_\text{SD}, \\
\text{s.t.}\quad  E_\text{D} \ge E_\text{H} \ge E_\text{SD};
&\quad c_i\in \mathcal{C}, \forall i \in \{1,2,\ldots,m\}.
    \label{eq:overall}
\end{split}
\end{equation}

Here, $E_\text{D}$, $E_\text{H}$, and $E_\text{SD}$ represent the objective values for discriminability, harmony, and spatial distribution, respectively.
$\mathcal{C}$ is the feasible color range.
The weighting parameters $\alpha$ and $\beta$ control the trade-offs between multiple objectives, which will be automatically determined during the optimization process.

\subsubsection{Color Discriminability}
\label{sec:discriminability}
Following Palettailor~\cite{lu2021palettailor}, the total objective function of color discriminability $E_{\mathrm{D}}$ consists of two terms: perceptual difference $E_{\mathrm{PD}}$ and name difference $E_{\mathrm{ND}}$.

\noindent\textbf{Perceptual difference}.
\label{sec:perceptual}
Perceptual difference quantifies the human-perceived difference between two colors.
In our implementation, we use the CIEDE2000 formula to calculate this perceptual difference because it is closely aligned with human perception~\cite{sharma2005ciede2000}.
Accordingly, the perceptual difference is defined as:

\begin{equation}
    E_\mathrm{PD} = \min_{1\le i< j\le n}D(c_i, c_j) + \min (\min_{1\le i< j\le n}D(c_i, c_j)-10, 0),
\end{equation}
where $D(c_i,c_j)$ is the perceptual difference between colors $c_i$ and $c_j$ using the CIEDE2000 formula.
The first term aims to maximize the minimal perceptual difference among all color pairs.
The second term introduces an extra penalty when the minimal perceptual difference falls below a threshold of 10, which is required to achieve high accuracy in judging whether two colors are identical.

\noindent\textbf{Name difference}.
In practice, colors that are perceptually different may still be described using the same name.
For example, these two colors \crule{20, 104, 255} and \crule{0,30,225} are both commonly described as ``Blue.''
Such naming ambiguity should be avoided since it leads to confusion when discussing colors in visualizations.
Heer and Stone~\cite{heer2012color} introduced the concept of name difference to quantify the likelihood that two colors are described using the same name.
They represented each color with a 153-dimensional feature vector, where each dimension corresponds to a popular color name.
Name difference is calculated by averaging the cosine distances between all color pairs:
\begin{equation}
    E_\mathrm{ND} = \frac{2}{m(m-1)} \sum_{1\le i< j\le n} (1-\cos \frac{T_{c_i}\cdot T_{c_j}}{\lVert T_{c_i}\rVert\cdot \lVert  T_{c_j}\rVert}),
\end{equation}
where $T_{c_i}$ and $T_{c_j}$ are the feature vectors of colors $c_i$ and $c_j$.

The color discriminability is then defined as a weighted combination of perceptual difference and name difference: $E_\mathrm{D} = \gamma_1 E_\mathrm{PD} + \gamma_2 E_\mathrm{ND}$. Following Palettailor~\cite{lu2021palettailor}, we set $\gamma_1=0.1$ and $\gamma_2=2.0$, which gives satisfactory results in practice.


\subsubsection{Color Harmony}
\label{sec:harmony}
We adopt the most advanced color harmony theory developed by Lara-Alvarez and Reyes~\cite{lara2019geometric}.
We choose this theory because it aligns well with human preferences and \vica{uses hue-chroma-lightness (CIELCh) color space, which is more uniform and thus suitable for research on optimizing colors.}
According to this theory, a harmonic color assignment result should follow specific patterns on both the hue wheel and the chroma-lightness plane.
Accordingly, the total objective function of color harmony $E_\text{H}$ consists of two terms: hue harmony $E_{\text{Hue}}$ and chroma-lightness harmony $E_{\text{CL}}$.

\begin{figure}[t]
\centering
\includegraphics[width=\linewidth]{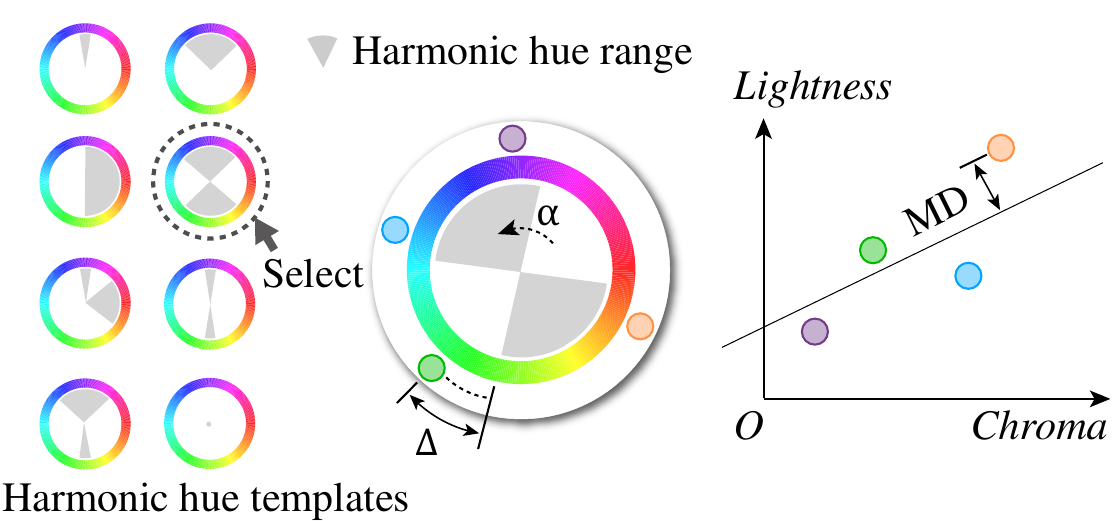}
\put(-200,-15){(a) Hue harmony}
\put(-75,-15){(b) C-L harmony}
\caption{
Color harmony: (a) on the hue wheel, colors that lie in the gray region are considered harmonic; (b) on the chroma-lightness plane, colors that follow a straight line are considered harmonic.
}
\vspace{-2mm}
\label{fig:hue}
\end{figure}

\noindent\textbf{Hue harmony}.
\vica{We do not directly use the hue harmony term proposed by Lara-Alvarez and Reyes~\cite{lara2019geometric} because it oversimplifies the widely-used Matsuda’s harmonic templates~\cite{matsuda1995color} defined on the hue-saturation-value (HSV) color space and directly applies them to the hue wheel of the CIELCh space.
The large difference between the hue wheels of these two spaces often leads to large discrepancies in color perception and harmony.
To tackle this issue, we use Matsuda’s templates to measure the hue harmony. 
Given a color assignment result, we first compute the corresponding hue values in the HSV space and then compare them with Matsuda’s templates.}
The difference between a color assignment result and a hue template is quantified by summing the minimal angular distances between the hue value of each color in the color assignment result and the hue range of the template, \vica{which can be freely rotated}.
Specifically, let $\theta_1,\cdots,\theta_m$ be the hue values of colors $c_1,\ldots,c_m$ in a color assignment result $C$, and $R$ be the hue range of a harmonic hue template, the hue difference between them is:

\begin{equation}
    \text{Hue\_Diff}(C, R) = \min_{\alpha\in [0, 360^\circ)} \sum_{i=1}^{m} \Delta(\alpha+\theta_i\ \mathrm{mod}\ 360^\circ, R),
\end{equation}
where $\alpha$ is the rotation angel of the hue range, and $\Delta(\theta, R)$ is the minimal angular distance between the value $\theta$ and the range $R$.
If $\theta$ falls within the range $R$, the minimal angular distance is 0. 
Otherwise, it is calculated as the distance between $\theta$ and the nearest boundary of $R$.
The objective value of hue harmony is then defined as the negative value of the smallest hue difference across all eight hue templates, and it is normalized to the range $[0,1]$ using min-max normalization.

\begin{equation}
    E_\mathrm{Hue} =  \text{Normalize}(- \min_{R}\ \ \text{Hue\_Diff}(C, R)) .
\end{equation}

\begin{figure*}[t]
    \centering
    \includegraphics[width=\linewidth]{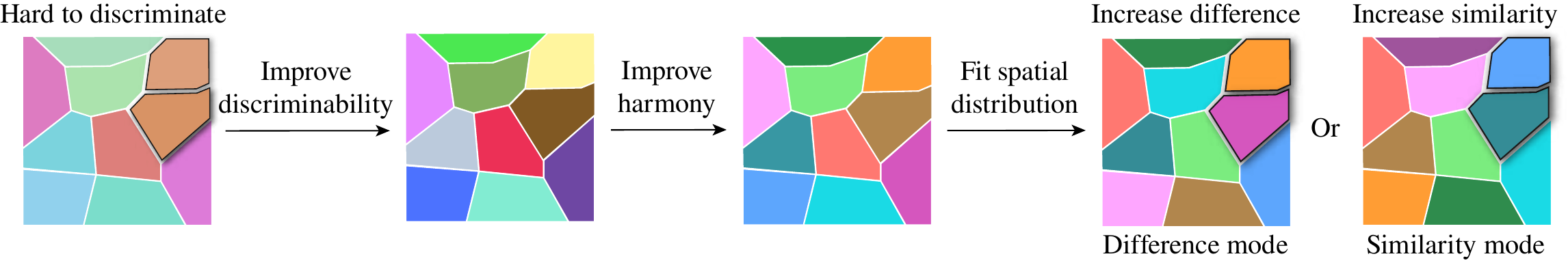}
    \caption{
    The continuation method improves assignment result by sequentially incorporating discriminability, harmony, and spatial distribution.
    }
      \vspace{-2mm}
    \label{fig:opt}
\end{figure*}
\noindent\textbf{Chroma-lightness harmony}.
\vica{We directly used the chroma-lightness harmony term proposed by Lara-Alvarez and Reyes~\cite{lara2019geometric}, which encourages colors to follow a straight line in the chroma-lightness plane}
As shown in Fig.~\ref{fig:hue}(b), given a color assignment result, it first determines the corresponding maximum likelihood line that best fits the colors in the chroma-lightness plane.
Next, it calculates the deviation of $i$-th color from this ideal line, denoted by $\mathrm{MD}_i$.
As it is unnecessary to strictly adhere to the line, we allow a deviation of 15 units on the chroma-lightness plane, which is recommended by Liu~\etal\cite{liu2022image}.
The optimization objective is then defined as:

\begin{equation}
    E_\text{LC}=
    \text{Normalize}(-\sum_{i=1}^n \max(\text{MD}_i - 15,0)).
\end{equation}

The color harmony is then defined as 
the sum of hue harmony and chroma-lightness harmony, which gives satisfactory results in practice.

\subsubsection{Spatial Distribution}
\label{sec:spatial}
When applying color assignment results in visualization, discriminability and harmony can be further enhanced by considering the spatial distribution of the visualized data.
For example, increasing the color differences between adjacent classes can enhance the color discriminability and make the boundaries clearer~\cite{Lee2013percept,lu2021palettailor}, and ensuring the harmony of color pairs between adjacent classes can generate more visually coherent results\cite{ou2006colour, ou2011additivity}.
Since the concept of adjacent classes changes with different types of visualizations, the optimization objective for data distribution is calculated by averaging the score of all neighboring sample pairs:

\begin{equation}
    E_{\text{SD}} = \frac{1}{|X|}\sum_{x_i\in X} \frac{1}{|\Omega_{x_i}|}\sum_{x_j\in \Omega_{x_i}}\frac{f(x_i, x_j)}{d(x_i, x_j)}.
\end{equation}
\begin{wrapfigure}{l}{0.111\textwidth}
\includegraphics[width=0.111\textwidth]{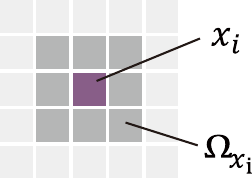}
\end{wrapfigure}
Here, $X$ is the set of all samples, and $\Omega_{x_i}$ is the set of neighboring samples of sample $x_i$ in visualization.
We consider three representative types of visualization: scatterplots~\cite{yang2020drift} (point-based), parallel coordinates~\cite{lu2021palettailor} (line-based), and grid visualizations~\cite{liu2017towards,chen2024unified} (area-based).
For grid visualizations, we consider eight surrounding cells of the center cell.
For scatterplots and parallel coordinates, we use eight nearest neighbors, which is consistent with grid visualizations.
The score of each sample pair is $f(x_i, x_j)/d(x_i, x_j)$.
Here, $d(x_i, x_j)$ is the spatial distance between samples $x_i$ and $x_j$ in the visualization.
Therefore, a closer pair would have more impact on the objective.
$f(x_i, x_j)$ is the optimization objective for each pair.
Based on the requirement analysis, we consider two modes for this term: difference mode and similarity mode.

The difference mode favors a color pair with a larger perceptual difference between adjacent classes.
Therefore, $f(x_i, x_j)$ is set as $D(c(x_i), c(x_j)) + P(c(x_i), c(x_j))$.
The first term is the perceptual difference between the colors of $x_i$ and $x_j$.
The second term measures the pair harmony between colors of $x_i$ and $x_j$.
We use the formula introduced in Ou's recent work about color pair harmony~\cite{ou2018universal} to calculate this term.

The similarity mode reduces the perceptual difference between similar classes to facilitate data exploration and understanding.
In our implementation,
$f(x_i, x_j)$ is defined as $-D(c(x_i), c(x_j)) \cdot s(x_i,x_j)
+ P(c(x_i), c(x_j))$, where $s$ measures the class similarity.
This similarity is determined by first averaging the feature vectors within each class to create class-level feature vectors and then calculating the similarities between them.
In the similarity mode, a larger perceptual difference for more similar class pairs results in a higher penalty.

\label{sec:opt}
\subsubsection{Optimization}

A straightforward way to solve the optimization problem defined in Eq.~(\ref{eq:overall}) is simulated annealing. 
It is chosen because of its flexibility to accommodate multiple objectives and its effectiveness in escaping local optima during the optimization process.
\vica{First, we use the blue noise sampling~\cite{cook1986stochastic} to generate the initial color assignment result within the default color range.
This technique ensures that colors are evenly distributed, which provides basic discriminability.}
At each iteration, the algorithm adjusts the color assignment result and re-evaluates the objective value.
Adjustments that improve the objective value are always accepted, while those that reduce the objective value are accepted with a progressively decreasing probability over time.
However, the simulated annealing algorithm suffers from slow convergence due to the low acceptance rate that comes with the original highly non-convex problem.
To accelerate convergence, we combine it with the continuation method.
Starting from solely optimizing discriminability, it sequentially incorporates harmony and spatial distribution.
Each solution to the previous problem serves as a starting point for the optimization of the subsequent problem.
By guiding the optimization process towards more promising regions in the solution space, this method achieves higher acceptance rates and faster convergence~\cite{bengio2009curriculum}.
The experimental result shows that our method can generate color assignment results for 30 classes in 1 second, which well supports real-time interaction for users navigating through hierarchical visualizations (see the supplemental material for more details).
When incorporating a new optimization goal in each stage of the continuation method, we dynamically set its weighting parameter using loss-balanced task weighting~\cite{liu2019loss,yang2024interactive}.
The basic idea is to ensure that different goals are optimized in a similar progress.
Thus, the weighting parameter of the goal with less progress will be increased so that it can be further improved in subsequent iterations.
Specifically, in each stage, the weighting parameter of the newly incorporated goal is set as the ratio between the current objective value and the possibly maximal value.
After the algorithm converges, the weighting parameter will be fixed, and the continuation method will move to the next stage.
Fig.~\ref{fig:opt} shows the incremental refinement of the color assignment results through each phase of the continuation method.
The idea of the continuation method also aligns with the typical process of hand-crafted color assignment design.
Initially, users select a set of distinguishable colors tailored to the number of classes.
Next, they adjust the hue, chroma, and lightness to improve harmony.
Finally, they assign these colors to different classes in the visualization and make slight adjustments to refine the overall visual effects.

\subsection{Color Range Selection}
When generating the color assignment result for classes at the top level, we use the default color range during the optimization process.
When generating the color assignment result for classes at the deeper level, we select an appropriate color range based on parent classes selected by users.
The colors of their child classes will be restricted within the selected color range.
This ensures that the generated color assignment result reflects the hierarchical structures within the data.

\subsubsection{Default Color Range}
Guided by expert interviews and the common practice in color assignment research~\cite{Gramazio2017colorgorical, tennekes2014treecolor, waldin2019cuttlefish, coombes2018polychrome}, we set the default range for chroma and lightness as $[40, 85]$ instead of the full range $[0, 100]$.
This excludes relatively extreme colors, including dim colors with low chroma/lightness (\eg, \crule{22,13,120}) or highly intense colors that are glaring (\eg, \crule{255,255,0}).
Moreover, some studies have pointed out that even within this range, there are still some strongly disliked colors~\cite{palmer2010ecological,yokosawa2010cross}, such as \crule{151,130,0} and \crule{98,112,38}.
To address this issue, we further exclude the range where \vica{lightness falls within $[40, 75]$ and hue simultaneously falls within $[85^\circ, 114^\circ]$,} as previously utilized by Gramazio~\etal~\cite{Gramazio2017colorgorical}.

\subsubsection{Dynamic Color Range}
\begin{figure}[t]
\centering
\includegraphics[width=\linewidth]{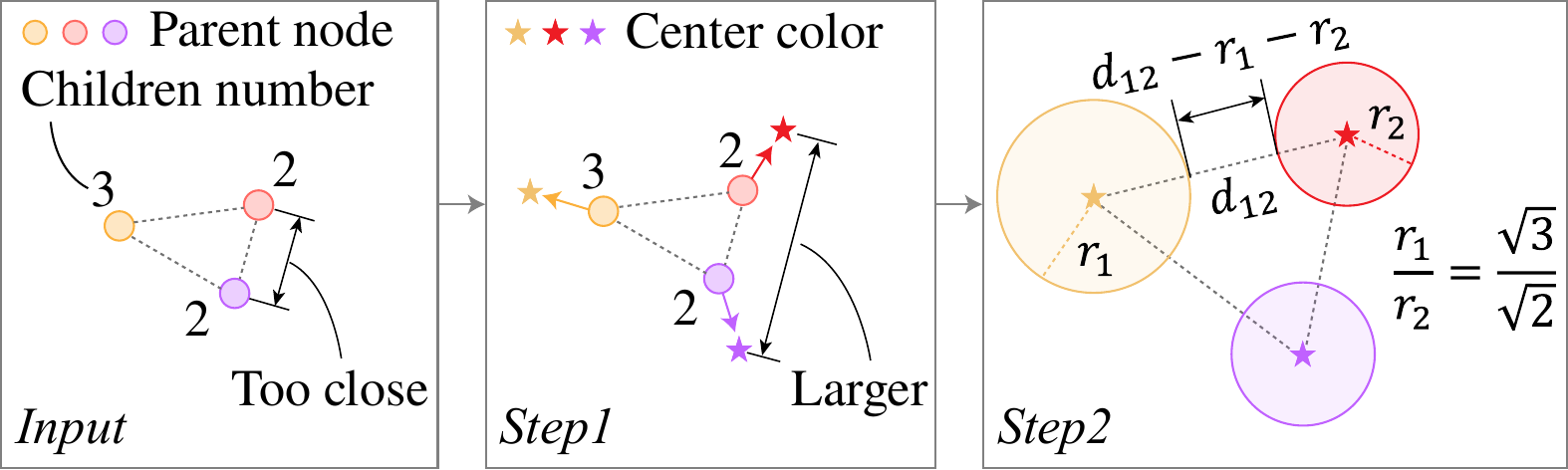}
\caption{The workflow of dynamic color range selection.}
\vspace{-4mm}
\label{fig:color-range-selection}
\end{figure}

For each parent class, we dynamically select a sphere based on the perceptual difference to guide the color assignment of its child classes.
\vica{We use spheres here because they allow us to ensure that the colors under the same parent are more similar than the colors under different parents by determining proper radii of spheres.}
However, when the colors of parent classes are close to each other, the available color range of each sphere becomes too narrow to generate high-quality color assignment results.
This issue becomes more common and more severe when users explore deeper hierarchical levels.
In practice, reusing the color range of invisible classes after zooming in can better exploit the color space \vica{and will not cause much confusion}~\cite{waldin2016chameleon,waldin2019cuttlefish}.
As shown in Fig.~\ref{fig:color-range-selection}, this can be achieved by first adjusting the centers of these spheres and then determining their radii.

\noindent\emph{Step 1: Adjust the centers of the spheres}.
We use the color assignment method introduced in Sec.~\ref{subsec:assignment} to adjust the centers of the spheres based on the colors of parent classes.
We choose it because of its effectiveness in ensuring discriminability and harmony.
However, directly applying this method can lead to two issues.
First, without a proper constraint on the adjustment range, colors may change excessively and thus increase the recognition burden.
Second, if the chroma or lightness of the colors of parent classes are close to the boundary of the default range ($[40, 85]$), it leaves less space to create aesthetically pleasing colors for child classes.
To address these two issues, we impose additional constraints during the simulated annealing process to adjust the sphere centers.
First, we ensure consistency between the initial colors and the adjusted colors.
Specifically, each adjusted color must remain closest to its initial color. 
By doing so, users can better maintain their mental map and correctly connect the initial colors with the adjusted colors.
Second, we narrow down the feasible range for chroma and lightness from $[40, 85]$ to $[45, 80]$.
This provides more opportunities to generate high-quality color assignment results for child classes.

\noindent\emph{Step 2: Determine the radii of the spheres}.
The goal of determining the radii is to ensure that the colors within the same sphere are closer than colors in different spheres and that the spheres with more child classes have larger radii to maintain discriminability.
Let $r_i$ and $r_j$ denote the radii of two spheres, and $d_{ij}$ denotes the distance between their centers.
The gap between these two spheres will be $d_{ij}-r_i-r_j$.
\begin{wrapfigure}{l}{0.12\textwidth}
\vspace{-5pt}
\includegraphics[width=0.12\textwidth]
{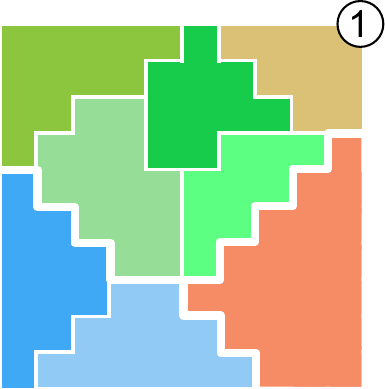}
\vspace{-12pt}
\end{wrapfigure}
First, we ensure that the gap must exceed the radii of both spheres, \ie, $d_{ij}-r_i-r_j > 1\max(r_i,r_j)$.
However, during the development of our methods, the experts pointed out that the hue plays a more important role in the identification of parent-child relationships, and only considering perceptual differences can sometimes result in misunderstanding.
For example, in the left image, the colors are sampled within the ranges of three spheres: green (top), blue (bottom-left), and orange (bottom-right).
Although the color on the top-right corner (\crule{227,194,64}) is closer to the green color than the orange and the blue, there is still a noticeable difference in its hue compared to the green color.
This potentially misleads users to perceive it as belonging to a separate parent class, \eg, a parent class with yellow color.
To avoid such misunderstanding, we add an additional restriction on hue in a similar way, which ensures that the gap between two hue ranges must exceed the length of both ranges.
Second, we study how to properly determine the radius of the sphere based on the number of child classes.
We conduct an experiment to estimate the relationship between the radius and the number of child classes through blue noise sampling~\cite{cook1986stochastic, yuan2021evaluation}.
Specifically, given a sphere with a radius $r$, we employ the widely adopted dart-throwing method~\cite{cook1986stochastic} to sample colors within it until no more discernible colors can be sampled.
Here, a discernible color means that the perceptual differences between it and those sampled colors exceed a threshold of 10, which is consistent with the threshold we used in Sec.~\ref{sec:perceptual}.
Our experimental results indicate a roughly linear relationship between the maximal number of possible colors $n$ and the square of the radius: $n\propto r^2$.
Therefore, we introduce constraints that $r_i/r_j=\sqrt{n_i}/\sqrt{n_j}$, where $r_i$ and $r_j$ denote the radii of two spheres associated with two parent classes, and $n_i$ and $n_j$ denote the number of child classes within those spheres.
Finally, the radii of the spheres are determined as the maximal radii that adhere to both $d_{12}-r_1-r_2 > \max(r_1,r_2)$ and $r_1/r_2=\sqrt{n_1}/\sqrt{n_2}$.

\begin{table*}[!t]
\fontsize{8}{8}\selectfont
\setlength{\tabcolsep}{1em}
\centering
\caption{Comparison of our method with the representative color assignment methods, with the best in \textbf{bold} and the second best \underline{underlined}. The values in \textcolor{Gray}{gray} indicate that the colors cannot be easily distinguished.}
PD: perceptual difference, ND: name difference, CL: chroma-lightness, BHDI: balanced harmony-discrimination index, SS: silhouette score, DR: distance ratio.
\vspace{-1mm}
\begin{tabular}{lcccccccccccc}
\toprule
\multirow{3}{*}[-1.2 ex]{Methods} & \multicolumn{5}{c}{Flat color assignment} & \multicolumn{7}{c}{Hierarchical color assignment} \\
\cmidrule(lr){2-6} \cmidrule(lr){7-13} 
& \multicolumn{2}{c}{Discriminability} & \multicolumn{2}{c}{Harmony} & \multirow{2}{*}{BHDI} & \multicolumn{2}{c}{Discriminability} & \multicolumn{2}{c}{Harmony}& \multirow{2}{*}{BHDI} & \multicolumn{2}{c}{Alignment with hierarchy} \\
\cmidrule(lr){2-3}\cmidrule(lr){4-5} \cmidrule(lr){7-8} \cmidrule(lr){9-10} \cmidrule(lr){12-13}
& PD & ND  & Hue & CL & & PD & ND & Hue & CL & & SS & DR \\
\midrule
Palettailor &19.419 & 0.913 & 0.296 & 0.377 & 4.441 & \textcolor{Gray}{8.202}  & 0.508 & 0.838 & 0.579 &3.253& 0.582 & 0.883  \\
Color Crafting &\textcolor{Gray}{2.549} & 0.261& \textbf{1.000}  & \textbf{1.000} & 2.777 & \textcolor{Gray}{4.176}  & 0.243 & \textbf{0.997}  & \underline{0.998}&2.898 & 0.642 & 0.930  \\
Tree Colors &\textcolor{Gray}{6.168} & 0.848 &0.608 & \textbf{1.000} & 3.909  & \textcolor{Gray}{0.226}  & 0.066 & 0.978 & \textbf{1.000} &2.133 & \textbf{0.785} & \textbf{0.994}  \\
Cuttlefish &\textcolor{Gray}{9.944} & 0.813 & 0.702 & \textbf{1.000} & 4.323 & \textcolor{Gray}{4.615} & 0.382 & 0.606 & \textbf{1.000} & 2.832 & 0.715 & 0.907  \\
\midrule
Ours-D &\textbf{23.194} & 
\textbf{0.921} & 
0.876 & 0.955 & \underline{5.992} & \textbf{16.579}  & \textbf{0.736} & 0.984 & 0.810 &\underline{4.926}& \underline{0.740} & 0.945 \\
Ours-S &\underline{23.070}  &\underline{0.920}  & \underline{0.893}  &\underline{0.962} & \textbf{6.002} & \underline{16.482}  & \underline{0.699} & \underline{0.985} & 0.927 &\textbf{4.958}& \underline{0.740} & \underline{0.946}  \\
\bottomrule
\end{tabular}
\label{table:res}
\end{table*}

\section{Evaluation}
\subsection{Quantitative Evaluation}

\noindent\textbf{Datasets}.
We evaluated the quality of different color assignment results across 12 datasets, which have been widely used in recent visualization research~\cite{xiang2019interactive, zhou2023cluster}.
Six of them (MNIST\cite{dataset-mnist}, Animals\cite{dataset-animals}, Indian Food\cite{dataset-food}, Isolet\cite{dataset-isolet}, Texture\cite{dataset-texture}, Clothing\cite{xiang2019interactive}) are flat datasets with a moderate number of classes, ranging from 10 to 26. 
These datasets are used to evaluate different flat color assignment methods.
The remaining six datasets (Food101~\cite{bossard2014food}, Flowers102~\cite{nilsback2008automated}, Stanford Cars~\cite{krause20133d}, Caltech256~\cite{griffin2007caltech}, NABirds~\cite{Horn_2015_CVPR}, ImageNet1k~\cite{deng2009imagenet}) have a larger number of classes, ranging from 101 to 1000.
These datasets are used to evaluate different hierarchical color assignment methods.
For datasets with a pre-existing class hierarchy, such as ImageNet, we directly used their hierarchies.
For the other datasets, we applied the commonly used hierarchical k-means method to build the class hierarchy in a top-down manner. 
More details of these datasets are given in the supplemental material.

We used these datasets to create three types of visualization: scatterplots (point-based)~\cite{xiang2019interactive,chen2021interactive}, parallel coordinates (line-based)~\cite{fua1999hierarchical}, and grid visualizations (area-based)~\cite{zhou2023cluster, chen2020oodanalyzer}, which cover the primary types of visualization techniques for data analysis.
We also evaluated the generated color assignment results on their own, \ie, considering the color palettes solely without integrating them into specific visualizations. 
When evaluating the flat color assignment methods, we assigned colors to all classes.
When evaluating hierarchical color assignment methods, 
we simulated how users explore the hierarchy by randomly choosing a subtree for expansion.

\noindent\textbf{Baseline methods and our method variations}.
We chose four state-of-the-art color assignment methods for comparison.
Palettailor~\cite{lu2021palettailor} and Color Crafting~\cite{smart2019color} are two representative flat color assignment methods.
Palettailor focuses on optimizing color discrimination and incorporates spatial distribution to improve this aspect further, while Color Crafting ensures basic color discrimination and focuses more on color harmony.
Tree Colors~\cite{tennekes2014treecolor} and Cuttlefish~\cite{waldin2019cuttlefish} are two representative hierarchical color assignment methods.
Tree Colors is a static method that assigns colors to all classes, while Cuttlefish is a dynamic method that only assigns colors to visible classes during exploration.
Since Palettailor and Color Crafting are flat color assignment methods, we extended them to support hierarchical color assignment.
Palettailor is capable of generating color assignment results within a specified hue range.
Therefore, we employed our dynamic color selection method to select the hue ranges for child classes and then used Palettailor to generate color assignment results within each range.
Color Crafting generates a sequence of colors with a similar hue but different lightness levels.
We first chose colors with different hues for the classes at the top level.
Subsequently, we estimated the lightness range for child classes and used Color Crafting to generate color assignment results for each child class within these specified lightness ranges.

We compared these methods with two modes of our method.
Ours-D is the difference mode, which increases the perceptual difference between colors of spatially adjacent classes.
Ours-S is the similarity mode, which reduces the perceptual difference between colors of similar classes.

\noindent\textbf{Evaluation criteria}.
We used seven measures to evaluate the quality of color assignment results from three perspectives: discriminability, harmony, and alignment with hierarchical structures.
Discriminability is evaluated using perceptual difference (PD) and name difference (ND), which have been introduced in Sec.~\ref{sec:discriminability}.
Harmony is measured using hue harmony (Hue) and chroma-lightness harmony (CL), which have been introduced in Sec.~\ref{sec:harmony}.
While discriminability and harmony are both crucial, they often conflict with each other.
For example, using identical colors maximizes harmony, but minimizes discriminability. 
To address this, we design a new measure, the balanced harmony-discrimination index (BHDI), to combine these four measures and give a more comprehensive one.
Specifically, we follow the hyperparameters we set in Sec.~\ref{sec:assignment} to combine them, \ie, $0.1\times\text{PD}+2.0\times\text{ND}+\text{Hue}+\text{CL}$.
Alignment with hierarchical structures is evaluated using silhouette score (SS) and distance ratio (DR), \vica{which are not used in our optimization process}.
SS~\cite{rousseeuw1987silhouettes} measures the compactness and separability of colors among child classes.
A higher value indicates that child classes of the same parent class have more similar colors, and those of different parent classes have more distinct colors.
DR quantifies color similarity between child classes and their parent classes.
It calculates the color distance from a child class to the closest class at the parent level and compares it to the distance to its actual parent class.
A ratio closer to 1 indicates a better alignment between child and parent classes.
\vica{A more thorough comparison with additional measures is available in the supplemental material.}

\begin{figure}[t]
\centering
\includegraphics[width=\linewidth]{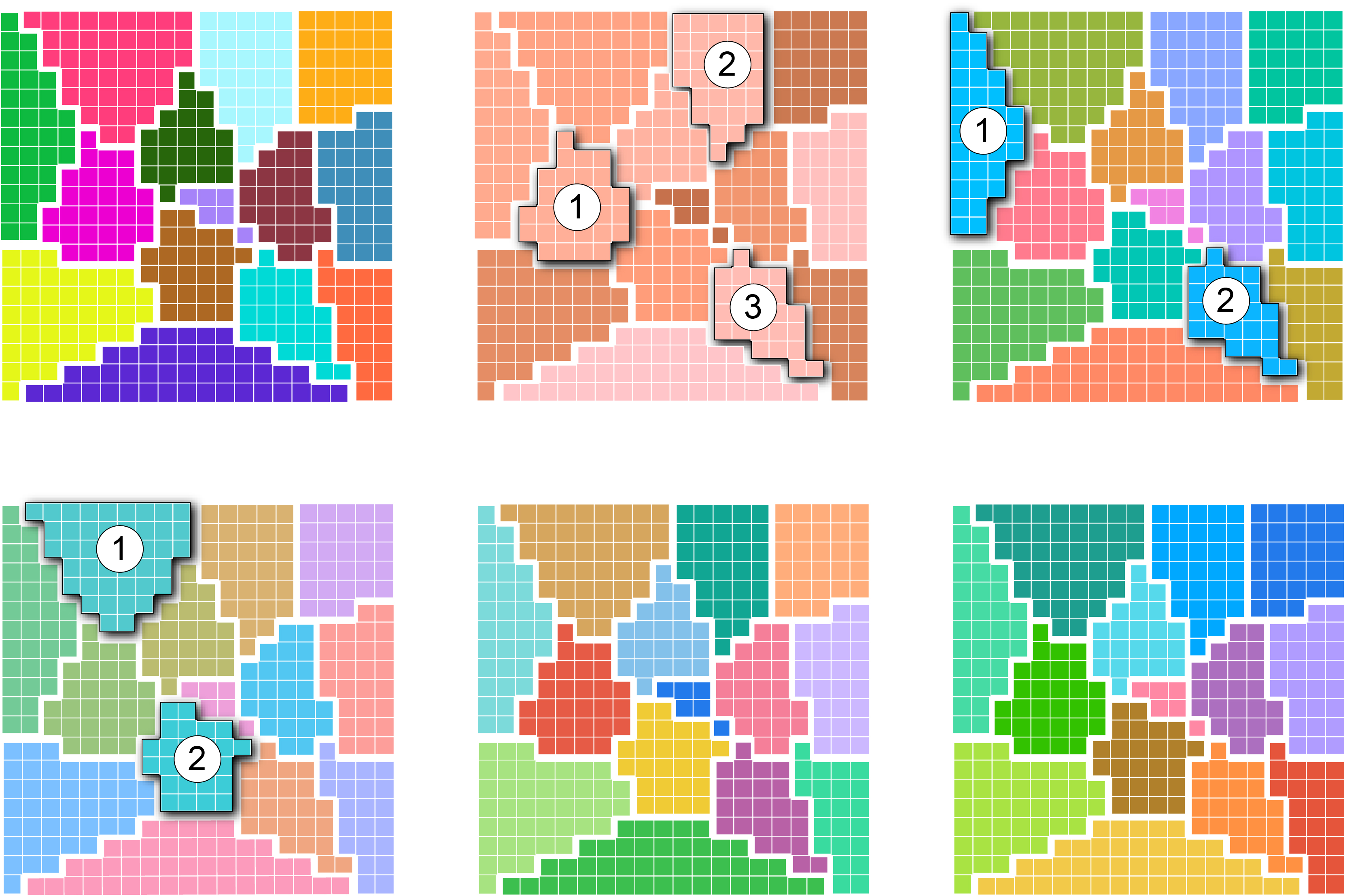}
\put(-240,82){(a) Palettailor}
\put(-157,82){(b) Color Crafting}
\put(-65,82){(c) Tree Colors}
\put(-238,-10){(d) Cuttlefish}
\put(-145,-10){(e) Ours-D}
\put(-55,-10){(f) Ours-S}
 \caption{Flat color assignment results generated by different methods.}
\label{fig:clothes}
\vspace{-4mm}
\end{figure}

\noindent\textbf{Results}.
Table~\ref{table:res} presents the comparison results between the baseline methods and our methods.
From the flat color assignment results, it can be seen that our methods perform best in terms of discriminability.
In particular, our methods achieve a perceptual difference of 23, which significantly exceeds the threshold of 10 for high accuracy in color discrimination~\cite{brychtova2017effect}.
In terms of harmony, our methods rank second in hue harmony, with only Color Crafting ahead.
However, it should be noted that Color Crafting's high hue harmony comes at the sacrifice of discriminability, which is evidenced by its worst discriminability (2.549 in perceptual difference and 0.261 in name difference) and the visualization results (\eg, colors \crule{242, 177, 154}\crule{243, 187, 172}\crule{244, 191, 181} in Fig.~\ref{fig:clothes}(b)).
Our methods fall behind in terms of chroma-lightness harmony compared to Color Crafting, Tree Colors, and Cuttlefish.
The main reason is that these methods apply either a fixed template or a strict constraint to determine the chroma and lightness of colors.
This results in worse discriminability, such as the colors \crule{89, 190, 251}\crule{85, 180, 250} in Fig.~\ref{fig:clothes}(c) and the colors \crule{81,201,205}\crule{60,204,215} in Fig.~\ref{fig:clothes}(d).
Instead, our methods sacrifice a little chroma-lightness harmony to allow greater variation to enhance discriminability.
Despite this, our methods perform best in terms of BHDI and produce visually appealing results (Figs.~\ref{fig:clothes}(e) and (f)).
These demonstrate the strength of our methods in balancing discriminability and harmony, which are both indispensable in visualization.
This is further confirmed by the user study results in Sec.~\ref{subsec:userstudy} as our methods are highly favored by the experts.

\begin{figure*}[t]
    \centering\includegraphics[width=\linewidth]{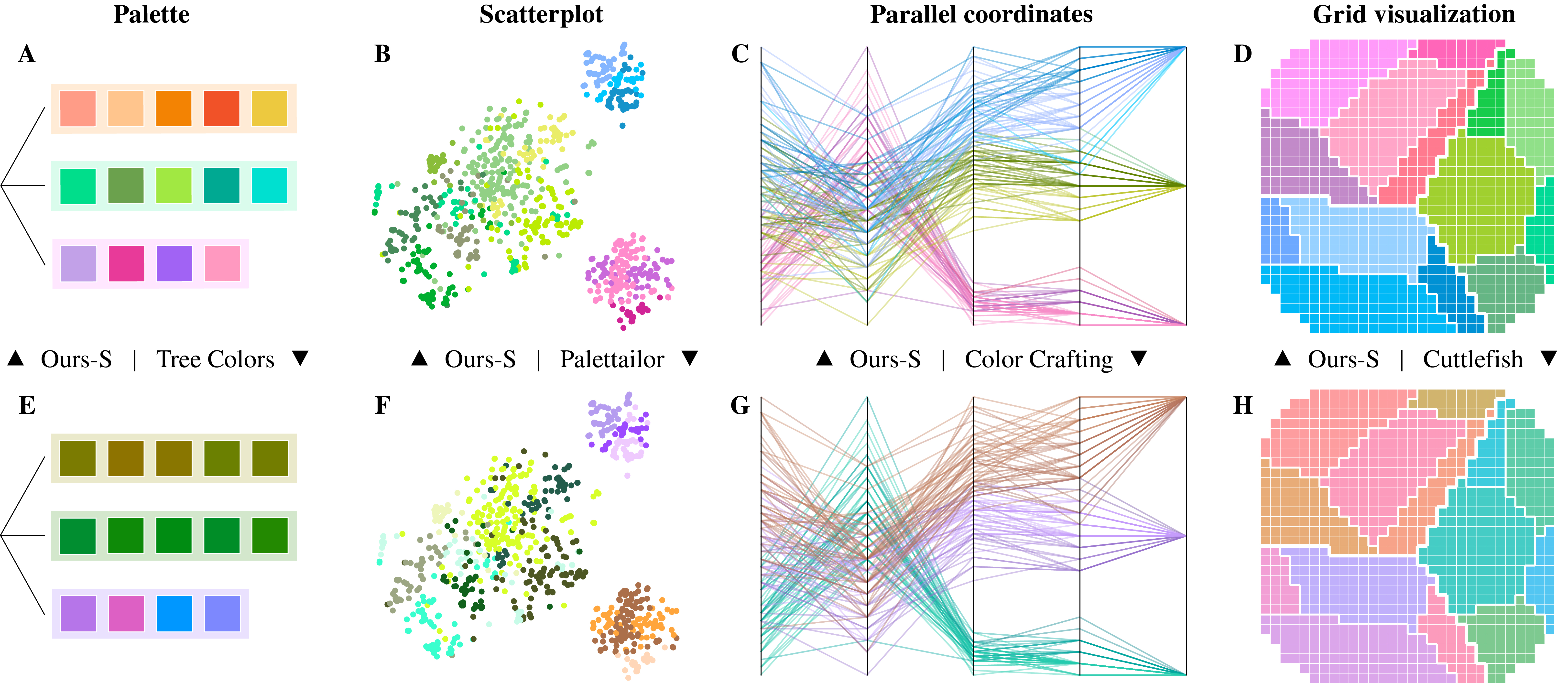}
    \caption{Hierarchical color assignment results generated by different methods and displayed in different visualization types.}
    \vspace{-4mm}
    \label{fig:hiererchy}
\end{figure*}

For hierarchical color assignment, the results on discriminability, harmony, and BHDI are similar to those for flat color assignment.
Therefore, we focus on how well these methods align with hierarchical structures.
As shown in Table~\ref{table:res}, our methods rank second among all methods, slightly behind Tree Colors.
The primary reason for Tree Colors' high performance is that it enforces the colors of child classes within a small range around the colors of parent classes, which achieves a tight alignment with hierarchical structures.
However, this results in extremely poor discriminability between colors, which is even worse than Color Crafting (0.226 \textit{vs.} 4.176 in perceptual difference and 0.066 \textit{vs.} 0.243 in name difference).
The visualization results in Fig.~\ref{fig:hiererchy}E also show that Tree Colors generates almost identical colors for five child classes of the same parent (\crule{65, 139, 61}\crule{65, 135, 38}\crule{64, 136, 42}\crule{65, 138, 54}\crule{71, 134, 35}).
In contrast, our methods generate colors that are both discriminable and clearly identifiable as belonging to the same parent class for these five child classes (\crule{105, 216, 146}\crule{120, 157, 86}\crule{177, 228, 92}\crule{79, 167, 147}\crule{0,224,208} in Fig.~\ref{fig:hiererchy}A).\looseness=-1

\begin{figure}[!t]
\centering
\includegraphics[width=\linewidth]{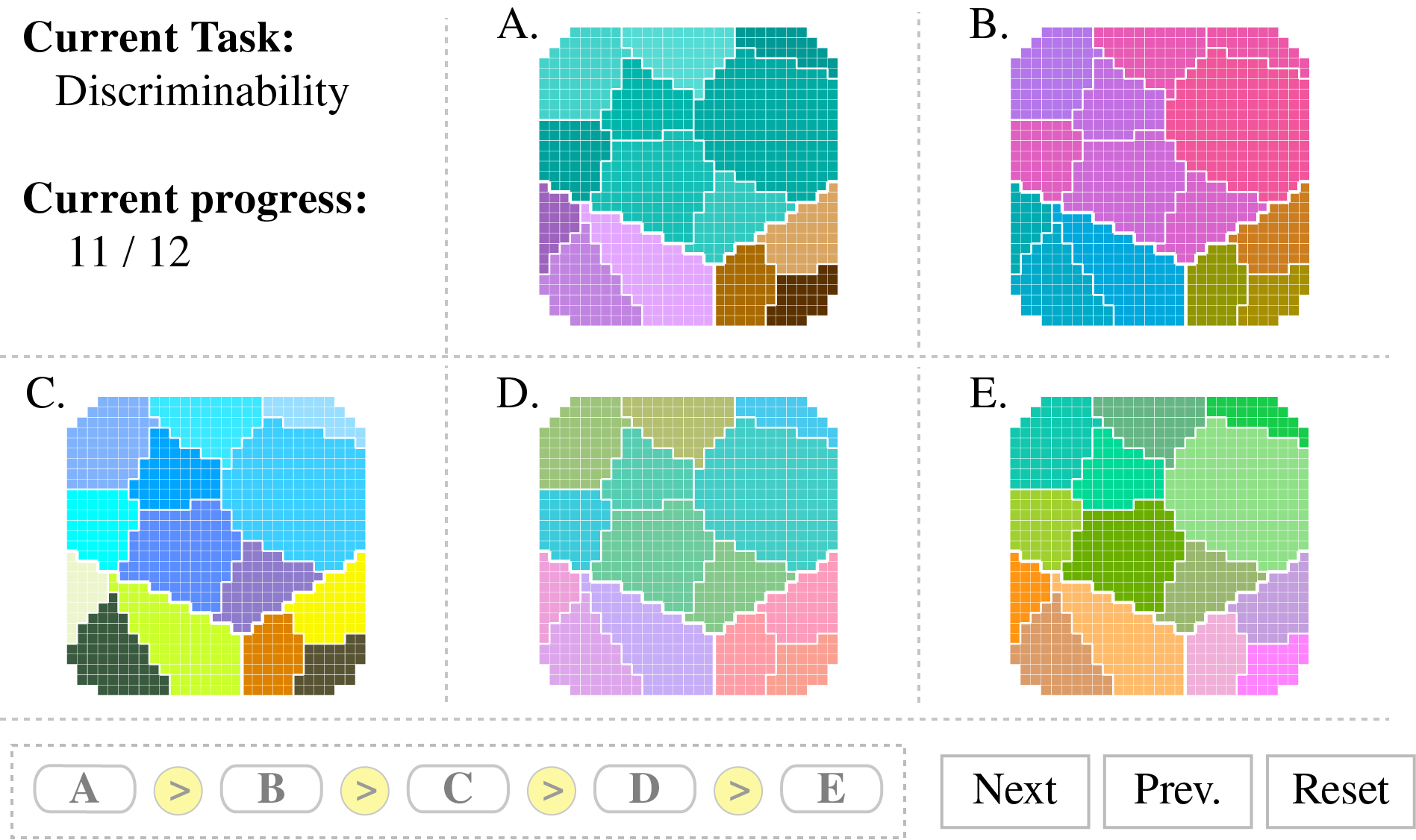}
\caption{The interface of the user study.}
\vspace{-4mm}
\label{fig:interface}
\end{figure}

\subsection{User Study}
We also conducted a user study to capture human preferences for different color assignment methods.

\subsubsection{User Study Design}
\label{subsec:userstudy}
\noindent\textbf{Methods}.
In our user study, we used the same four baseline methods as in our qualitative evaluation.
As shown in Table~\ref{table:res}, only minor differences are observed between the difference mode (Ours-D) and the similarity mode (Ours-S).
This prompts us to choose one to simplify the comparison process for experts. 
We chose the similarity mode because it shows slightly better performance in BHDI and alignment with hierarchical structures, and this mode is more suitable for data analysis in practice~\cite{xue2023reducing}.

\noindent\textbf{Experts}.
We recruited 20 experts for our study, including 12 males and 8 females.
Nine of them specialize in color design, and the remaining eleven are experts in information visualization.
All of them confirmed their expertise in color design and data visualization.
None of them reported any color deficiency.
Upon completion, each expert was rewarded with a \$30 gift card.

\noindent\textbf{Study procedure}.
The user study consists of three tasks that require the experts to rank the color assignment results based on discriminability, harmony, and alignment with hierarchical structures, respectively.
Before each task, we provided a brief overview of these concepts.
Following this, experts were required to complete 12 trials for each task.
In each trial, experts ranked the color assignment results on a web-based interface (Fig.~\ref{fig:interface}).
They could take a brief break after completing each task.
Upon completing all trials, they were asked to complete a questionnaire, which collected their personal information and detailed feedback on how they evaluated the five color assignment results depicted in Fig.~\ref{fig:interface}.
Each study lasted 40-60 minutes.

\noindent\textbf{Conditions and design}.
Similar to the quantitative evaluation, we included palettes, scatterplots, parallel coordinates, and grid visualizations in our user study to cover different types of visualization.
This diverse inclusion ensures that our evaluation results are more robust and reliable.
However, to avoid overburdening experts, we did not include all the 12 datasets used in the quantitative evaluation, as doing so would result in a total of 144 trials (3 tasks $\times$ 4 visualizations $\times$ 12 datasets).
This would require an excessive amount of time from our experts.
Instead, we considered three representative scenarios in data exploration:
1) examining a higher hierarchical level with a balanced subclass distribution (high-bal), which is the most common case in data exploration;
2) examining a higher hierarchical level with an imbalanced subclass distribution (high-imbal), which sometimes happens during exploration;
3) examining a lower hierarchical level (low), where the available color range will become much narrower.
This makes our findings more convincing and applicable to a wide range of data exploration scenarios.
Consequently, each expert went through 36 trials (3 tasks $\times$ 4 visualizations $\times$ 3 scenarios).

\subsubsection{Result Analysis}

\begin{figure*}[t]
    \centering
    \includegraphics[width=\linewidth]{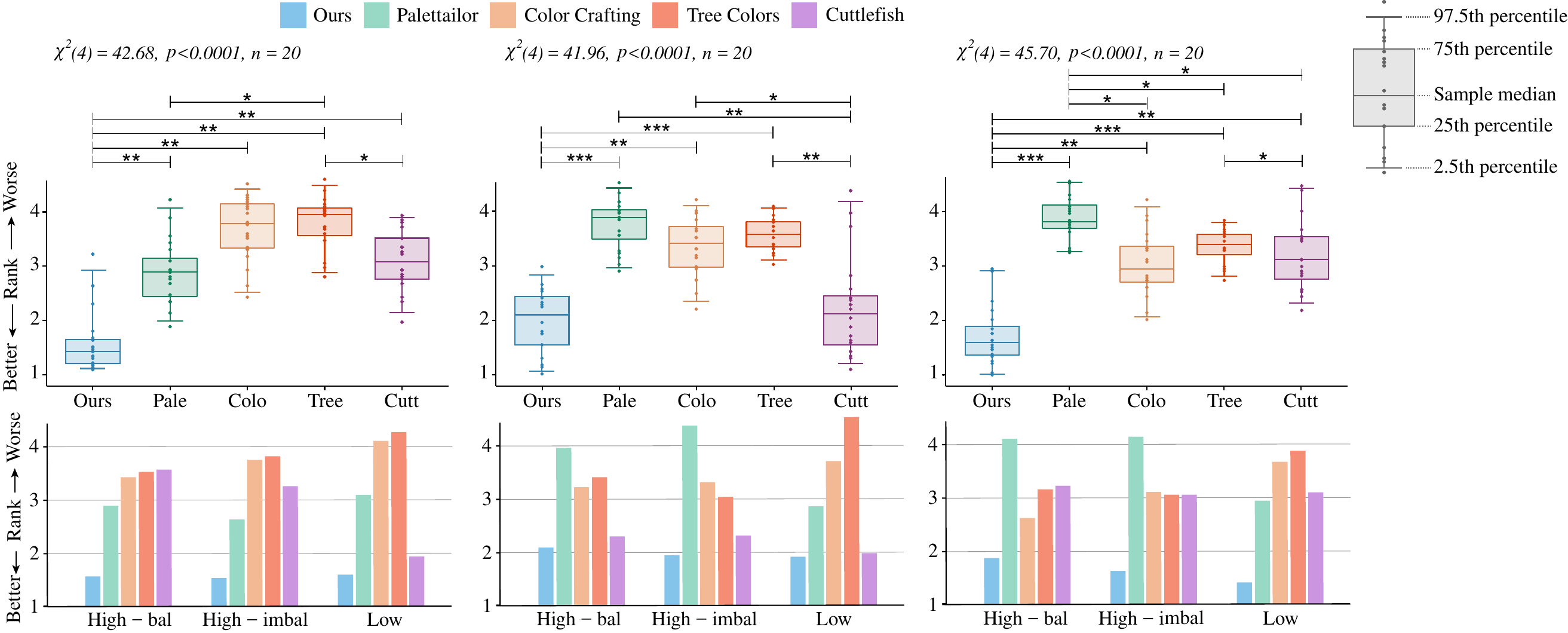}
    \put(-468,-10){(a) Discriminability}
    \put(-312,-10){(b) Harmony}
    \put(-218,-10){(c) Alignment with hierarchical structures}
    \caption{User study results in three tasks. Top: Friedman tests and pairwise Wilcoxon signed-rank tests on five methods.
    Bottom: Detailed comparison of their rankings under three exploration scenarios.
    * indicates $p<0.05$, ** indicates $p<0.01$, and *** indicates $p<0.001$.}
    \label{fig:userstudy}
    \vspace{-4mm}
\end{figure*}

\noindent\textbf{Overall comparison}.
First, we computed the average rank for each color assignment method.
If an expert ranked two methods equally, their ranks were set as their average rank.
For example, if the ranking is A=B>C=D=E, the ranks of A and B will be (1+2)/2=1.5, and the ranks of C, D, and E will be (3+4+5)/3=4.
Next, we conducted Friedman tests and pairwise Wilcoxon signed-rank tests to compare the ranks of different methods.
The statistical test results and the box plots are presented in Fig.~\ref{fig:userstudy}.
The Friedman test results indicate significant differences among the methods in discriminability ($\chi^2(4)=42.68,p<0.0001$), harmony ($\chi^2(4)=41.96,p<0.0001$), and alignment with hierarchical structures ($\chi^2(4)=45.70,p<0.0001$).
Therefore, we focus on pairwise comparisons in subsequent analysis.
Regarding discriminability, our method achieves an average rank of 1.64, which is significantly better than all the baseline methods.
Regarding harmony, our method achieves a comparable performance with Cuttlefish (an average rank of 2.04 \textit{vs.} 2.19), and they are both significantly better than the other three methods.
Regarding the alignment with hierarchical structures, our method ranks highest on average (1.70), which is significantly better than other methods.
These findings are consistent with the ones in the quantitative evaluation that our method is effective in balancing discriminability and harmony, while also maintaining better alignment with hierarchical structures.
Further details of the distribution of the ranks are summarized in the supplemental material.

\noindent\textbf{Detailed comparison on different scenarios}.
We further investigate whether the ranking results vary across different exploration scenarios.
As shown in Fig.~\ref{fig:userstudy}, the average ranks of different methods do not significantly change between the scenario of exploring a higher hierarchical level with a balanced subclass distribution and that with an imbalanced class distribution (high-bal \textit{vs.} high-imbal).
However, in the exploration of lower hierarchical levels, both Color Crafting and Tree Colors exhibit a notable drop in discriminability.
This is because, at lower levels, they produce color assignment results within a narrower color range.
Moreover, the experts ranked these two methods lower in harmony.
This is because they believed harmony is built on contrast, colors that are similar but not identical are perceived as more harmonic.
In contrast, both our method and Cuttlefish maintain good performance in discriminability and harmony, which is achieved by dynamically expanding the color range in lower levels.
Notably, our method achieves better alignment with hierarchical structure compared to Cuttlefish because of the improved dynamic color range selection method.
This improvement ensures a closer correspondence between the colors assigned to child classes and their respective parent classes, thereby improving overall hierarchical coherence.

\section{Expert Feedback and Discussion}
After the user study, we summarized expert feedback on their evaluation of different color assignment methods in terms of discriminability, harmony, and alignment with hierarchical structures.
To gain deeper insights, we conducted semi-structured interviews with eight experts.
In each interview, we first presented their ranking results and encouraged them to explain their choices, especially those where their opinions differed from the majority.
After going through all the results, we held an open discussion to collect more feedback on our color assignment method and explore potential opportunities for enhancement.
The duration of each discussion was around 50-60 minutes.

\noindent\textbf{Discriminability}.
When discussing how they compare different color assignment results in terms of discriminability, many experts highlighted the important role of hue and saturation.
This led to a higher ranking for options C/D/E compared to A/B in Fig.~\ref{fig:interface}.
Two experts also pointed out that to develop a color assignment method for a
wide audience, it is necessary to consider users with color vision deficiency (CVD).
Given the flexibility of our method, we are optimistic about its potential to generate CVD-friendly color assignment results.
The key lies in selecting an appropriate color range and refining the calculation of color discriminability.
Some research on CVD models~\cite{brettel1997computerized,machado2009physiologically} can be integrated to achieve this goal.
\vica{In addition, Stone~\etal\cite{stone2014engineering}
explored the impact of mark size on discriminability. 
A promising research direction involves incorporating 
the findings into 
the optimization process to tailor color assignment results for various visualization types, including scatterplots, line charts, and grid visualizations. 
\looseness=-1}

\noindent\textbf{Harmony}.
Many experts pointed out that they would first rank color assignment results with extremely high saturation or luminance, such as Fig.~\ref{fig:hiererchy}F and Fig.~\ref{fig:interface}C, as the least harmonic.
They also noted that harmony decreased when colors were too similar to be easily distinguished~\cite{ou2018universal}.
This explains the lower ranking of Color Crafting and Tree Colors in lower hierarchical levels.
Although our method and Cuttlefish scored lower in the harmony metric, they were still preferred by experts.
Moreover, several experts pointed out that personal preferences and cultural differences also affect the perception of color harmony~\cite{o2010colour,palmer2010ecological,yokosawa2010cross}.
For example, one expert favored warm colors with low saturation, such as the results generated by Cuttlefish (Fig.~\ref{fig:hiererchy}H).
Our methods can integrate these individual and cultural preferences by modifying the color ranges or including customized harmonic patterns.
However, it is difficult for end users to accurately describe their preferences in terms of color ranges.
It remains an opportunity to explore how to collect user feedback (like/dislike) and model user preferences.

\noindent\textbf{Consistency with hierarchical structures}.
The experts noted that when comparing different methods, they focused more on the perceptual difference between the colors of parent classes and child classes.
A smaller difference indicates a clearer parent-child relationship.
They also highlighted the critical role of hue difference between different parent classes, which would significantly facilitate the accurate identification of parent-child relationships.
This justifies our additional emphasis on the hue channel in the dynamic color range selection process.
Moreover, some experts prioritized discriminability between the colors of parent classes and would tolerate relatively smaller discriminability between the colors of child classes of the same parent.
Other experts held the opposite view and believed that maintaining discriminability between the colors of child classes was more important.
Therefore, it is necessary to provide a customizable trade-off between these two goals.
Currently, our method ensures that the gap between two color ranges for the child classes of different parents should be greater than the radii of each color range.
Users can modify this requirement to achieve a smaller or larger gap to adapt to their needs.

\section{Conclusion}
We develop a dynamic color assignment method that simultaneously considers discriminability, harmony, and spatial distribution.
It also dynamically assigns colors based on user exploration and aligns them with hierarchical structures.
Our method starts by generating a discriminable and harmonic color assignment result for top-level classes within the full color range.
When users zoom in on a region for detailed analysis, our method selects an appropriate color range for the child classes based on the colors of the selected parent classes.
Subsequently, our method generates the color assignment result for child classes within the selected color ranges and ensures discriminability and harmony.
The effectiveness of our method is demonstrated through a quantitative evaluation and a user study, which highlights its capability in generating high-quality dynamic color assignment results.
\
\section*{Supplemental Materials}
All supplemental materials are available on OSF at \url{https://osf.io/e4b5u/?view_only=68cc67c194c443b498bd2545ef551faa}, released under a CC BY 4.0 license.
In particular, they include (1) dataset information,  (2) the running time of our color assignment methods, (3) additional color assignment results, (4) additional quantitative evaluation results, (5) additional user study results, and (6) the video.

\acknowledgments{
This work was supported by the National Natural Science Foundation of China under grants U21A20469, 61936002, grants from the Institute Guo Qiang, THUIBCS, and BLBCI. The authors would like to thank Jun Yuan, Zhen Li, and Duan Li for their valuable contributions to the discussions, and Yiwei Hou for her assistance in voicing our video.
}

\bibliographystyle{abbrv-doi-hyperref}

\bibliography{reference}


\end{document}